\documentstyle[twoside,fleqn,espcrc2]{article}
\newcommand{\be}{\begin{equation}}
\newcommand{\ee}{\end{equation}}
\newcommand{\bra}[1]{\mbox{$\langle\, #1 \mid$}}
\newcommand{\ket}[1]{\mbox{$\mid #1\,\rangle$}}

\newcommand{\expec}[1]{\mbox{$\langle\, #1\,\rangle$}}
\newcommand{\expecl}[1]{\mbox{$\left\langle\, 
            \strut\displaystyle{#1}\,\right\rangle$}}

\title{Semiclassical Collapse of a Sphere of Dust and Hawking Radiation}
\author{Roberto Casadio\address{Dipartimento di Fisica, Universit\`a di
Bologna and\\ 
Istituto Nazionale di Fisica Nucleare, Sezione di Bologna, Italy \\
e-mail: casadio@bo.infn.it}}
\begin{document}
\begin{abstract}
The independent dynamical variables
of a collapsing homogeneous sphere of dust
are canonically quantised and coupled equations
describing matter (dust) and gravitation are obtained.
The conditions for the validity of the adiabatic 
(Born--Oppenheimer) and semiclassical approximations
are derived.
On neglecting back--reaction effects, it is also shown
that in the vicinity of the horizon and inside the dust
the Wightman function for a conformal scalar field
coupled to a monopole emitter is thermal at the
characteristic Hawking temperature.
\end{abstract}
\maketitle
\section{The semiclassical collapse}
\label{semicl}
The total classical action for a self gravitating sphere of homogeneous 
dust immersed in the vacuum (Schwarzschild) space--time is taken to be 
a generalization of the Oppenheimer--Snyder model \cite{oppenheimer,cv} 
($c=K_{Boltzmann}=1$, $\kappa\equiv 8\,\pi\,G={\ell_p}^2/\hbar$):
\begin{eqnarray}
S&=&-\int {d\tau\over2\,\kappa}\,\left(K_c\,{\dot K_c}^2+\epsilon\,K_c
\right)  \nonumber \\
&&
+\int {d\tau\over 2}\,K_c^3\,\left({\dot\phi}^2-{\mu}^2\,{\phi}^2\right)
\ ,
\label{s1}
\end{eqnarray}
where the dust is described by the scalar field $\phi=\phi(\tau)$ 
(to allow for change in the number of particles in the second quantized
development), $\mu=m/\hbar$ is the inverse of its Compton wave length and 
$K_c=K_0\,\partial_\eta h_\epsilon(\eta)$, with
\be
h_\epsilon(\eta)=\left\{
\begin{array}{ll}
\eta-\sin\eta\ \ &\ \ \epsilon=+1 \\
\sinh\eta-\eta\ \ &\ \ \epsilon=-1 \\
\eta^3/6\ \ &\ \ \epsilon=0\ ,
\end{array}
\right.
\ee
is the conformal factor of the interior (Robertson--Walker) metric
with positive, negative and zero curvature respectively,
\begin{eqnarray}
ds^2&=&-d\tau^2+{K_c}^2\,\left[{d\rho^2\over 1-\epsilon\,\rho^2}+
\rho^2\,d\Omega^2\right] \\
&& 0\le\rho\le\rho_{_0} \nonumber 
\ .
\end{eqnarray}
The constant $K_0$ is related to the mass parameter $M$ of the
external Schwarzschild metric by the junction condition 
$M={\rho_{_0}}^3\,K_0$ 
at the surface of the sphere $r_{_0}\equiv\rho_{_0}\,K_c$ \cite{mtw}.
\par
Canonical quantisation leads to the Wheeler--DeWitt equation
for the total wave function $\Psi=\Psi(K,\phi)$ \cite{cv}:
\be
\left[{\ell_p^2\over\kappa}{\partial_K^2\over K}
-{\epsilon\over\kappa}\,K
-{\hbar^2\over K^3}\,
{\partial^2\over\partial\phi^2}
+\mu^2\,\phi^2\,K^3
\right]\Psi=0
.
\label{wdw}
\ee
One may now choose a suitable operator ordering in the 
gravitational kinetic term and, following a previously employed 
procedure \cite{brout} (analogous to the Born--Oppenheimer
approximation used in molecular dynamics),
express $\Psi$ in the factorized form
\be
\Psi(K,\phi)=K\,\tilde\psi(K)\,\tilde\chi(\phi,K)
\ ,
\label{factor}
\ee
where $\tilde\psi$ and $\tilde\chi$ satisfy the following
equations \cite{cv}:
\begin{eqnarray}
\left[\hat H_{_G}\,K+K\,\expec{\hat H_{_M}}
\right]\,\tilde\psi 
=-{\kappa\,\hbar^2\over 2}\,\expec{\partial_K^2}\,\tilde\psi
\ ,
\label{wdw_g}
\end{eqnarray}
\begin{eqnarray}
\left[\hat H_{_M}-\expec{\hat H_{_M}}\right]\,\tilde\chi
+{\kappa\,\hbar^2}\,
\left({\partial\tilde\psi\over\partial K}\right)\,
{\partial\tilde\chi\over\partial K}  \nonumber \\
={\kappa\,\hbar^2\over2}\,\tilde\psi\,\left[
\expecl{{\partial^2\over\partial K^2}}-\partial_K^2\right]\,
\tilde\chi
\ ,
\label{wdw_m}
\end{eqnarray}
and $\expec{\hat A}\equiv \int d\phi\,\tilde\chi^\ast(K,\phi)\,
\hat A\,\tilde\chi(K,\phi)$ for any operator $\hat A$.
\subsection{The dust wave function}
\label{m_w}
If we consider the semiclassical (WKB) approximation to the wave 
function $\tilde\psi$, one has
\be
{\partial\ln\tilde\psi\over\partial K}\simeq
-{i\over\hbar}\,{\partial S_{eff}\over\partial K}
=-{i\over\hbar}\,\pi_{_K}
\ ,
\label{wkb}
\ee
where $S_{eff}$ is the effective action satisfying
the Hamilton--Jacobi equation associated with the
L.H.S. of Eq.~(\ref{wdw_g}).
In such a semiclassical limit $\tilde\psi$ will be 
peaked at the classical trajectory $K_c$.
One may then define a (conformal) time variable
\be
{\partial\over\partial\eta}\equiv
-i\,\hbar\,\kappa\,{\partial\ln\tilde\psi\over\partial K}\,
{\partial\over\partial K}
\ .
\label{eta}
\ee
Further if the R.H.S. of Eq.~(\ref{wdw_m}) is small
one gets from Eq.~(\ref{wdw_m}) the Schr\"odinger 
equation for a harmonic oscillator
\be
K_c\,\hat H_{_M}\,\chi_s=i\,\hbar\,
{\partial\chi_s\over\partial\eta}
\ ,
\label{schro}
\ee
where we have scaled the dynamical phase, 
$\chi_s\equiv\tilde\chi\,\exp\left\{-{i\over\hbar}\,\int^\eta
\expec{\hat H_{_M}(\eta')}\,K_c\,d\eta'\right\}$.
In the adiabatic approximation for $K_c$ the solutions 
are \cite{cv}:
\be
\tilde\chi_{_N}=N_{_N}\,H_{_N}(\gamma\,\phi)\,
e^{-\gamma^2\,\phi^2/2}
\ ,
\label{chi_s}
\ee
where $H_{_N}$ is the Hermite polynomial of degree 
$N$, $N_{_N}$ a normalization factor and
${\gamma}^2\equiv\mu\,{K_c}^3/\hbar$.
The expectation value of the matter Hamiltonian on the solutions
is given by
\be
K_c\,\expec{\hat H_{_M}}_{_N}=(N+1/2)\,\hbar\,\mu\,K_c
\ .
\label{e_m}
\ee
\par
One can build a coherent superposition of the states in
Eq.~(\ref{chi_s}) corresponding to a minimum wave packet 
oscillating about $\phi=0$ with classical frequency $\hbar\,\mu\,K_c$,
which is constant in the adiabatic approximation,
thus leading to a classical trajectory for the dust in the classical 
limit ($\hbar\to 0$) \cite{cv}.
\subsection{The gravitational wave function}
\label{s_wkb}
It is straightforward to derive an expression for $\tilde\psi$
which satisfies Eq.~(\ref{wdw_g}) in the semiclassical limit 
with vanishing R.H.S. (fluctuations).
Suppose that at time $\eta_{_0}$ for which $\partial_\eta K_c=0$ 
the gravitational wave function is a gaussian packet centred
at the value $K_c(\eta_{_0})$ with width $b$.
At a succeeding time $\eta$, $\tilde\psi$ will be given by
\be
\tilde\psi(K,\eta)=\int dK'\,
G_\epsilon(K,\eta;K',\eta_{_0})\,
\tilde\psi(K';\eta_{_0})
\ ,
\ee
where $G_\epsilon$ is the Green's function for the (inverted) 
harmonic oscillator for $\epsilon=+1(-1)$ or for a particle moving 
in a linear potential for $\epsilon=0$.
One then obtains \cite{cv}
\be
|\tilde\psi(K,\eta)|^2=\exp\left\{-\alpha^2\,
\left(K-K_c(\eta)\right)^2\right\}
\ ,
\label{psi}
\ee
where
\be
\alpha={b\over\left[\kappa^2\,\hbar^2\,
(\partial^2_\eta h_\epsilon)^2+b^4\,
\left(1-\epsilon\,(\partial^2_\eta h_\epsilon)^2\right)
\right]^{1/2}} 
\ .
\label{alpha}
\ee
We further note that the total gravitational wave function 
$K\,\tilde\psi(K)$ is $0$ for $K=0$, thus giving no finite probability
to fall exactly into the point--like singularity $K_c=0$.
\par
If one now considers the limit $\hbar\to 0$ followed by $b\to 0$
({\em classical point--like limit\/})
the classical trajectory is obtained:
\be
\alpha\to\infty\ \ \ \Longrightarrow\ \ \ 
{K^2\,|\tilde\psi|^2\over\bra{\tilde\psi}\,\hat K^2\,\ket{\tilde\psi}}
\to\delta(K-K_c)
\ ,
\label{hp=0}
\ee
where $\bra{\tilde\psi}\,\hat K^2\,\ket{\tilde\psi}$ 
is the norm of the complete wave function $K\,\tilde\psi$. 
With the above limits interchanged, that is $b\to 0$ 
with $\hbar$ finite, one would obtain
\be
\alpha\to 0\ \ \ \Longrightarrow\ \ \ 
{K^2\,|\tilde\psi|^2\over\bra{\tilde\psi}\,\hat K^2\,\ket{\tilde\psi}}
\to 0
\ .
\label{b=0}
\ee
However, we expect that when $b$ becomes smaller than the Planck 
length $\ell_p$ quantum gravitational effects (fluctuations)
become significant.
Hence it is more sensible to consider the limit for which one
has $b\sim\ell_p$ (corresponding to a minimum size wave packet
of the order of the Planck length) and then consider $\ell_p\to 0$,
which again leads to Eq.~(\ref{hp=0}).
\par
From Eq.~(\ref{e_m}) and taking into account the junction condition 
at the surface $\rho_0$ one may conclude that the total mass 
$M_{_G}\equiv M/\kappa$ of the collapsing sphere is proportional to 
the number of dust quanta $N$,
\be
M_{_G}={\rho_{_0}}^3\,N\,\hbar\,\mu
\ .
\label{m=sum}
\ee
The above apparently differs from the result obtained previously,
$M_{_G}\sim m_p\sqrt{n}$ \cite{bekenstein}, 
but we observe that the $N$ appearing in Eq.~(\ref{m=sum})
is an energy quantum number of the dust, while the $n$ 
in Bekenstein's formula can be obtained in the present context
from quantising the gravitational Hamiltonian with purely classical 
matter source in the $\epsilon=+1$ case only \cite{cv}.
\subsection{Consistency conditions}
\label{test}
Even if $\tilde\psi(K,\eta)$ is peaked on the classical
value $K_c$, there are fluctuations around it
and one must require Ehrenfest's theorem
to hold in order to allow for the junction condition.
In our case this means that one must have
\be
\Delta\equiv|\expec{\hat K^2}-\expec{\hat K}^2|/{\hat K}^2\ll 1
\ .
\label{delta}
\ee
For the wave function $\tilde\psi$ in Eq.~(\ref{psi}) one finds
$\Delta\sim{1/\alpha^2}\to 0$
in the classical point--like limit, Eq.~(\ref{hp=0}).
However if one considers the limiting procedure used in
Eq.~(\ref{b=0}) one has $\Delta\simeq 0.18$.
\par
It is particularly interesting to consider small black holes.
Let us then take $K_c$ small at fixed $\hbar$ and 
$b\sim\ell_p$, one then has $\Delta\sim \ell_p/K_c$,
which implies that one must have 
$r_{_0}(\eta)=\rho_{_0}\,K_c(\eta)\gg\ell_p$,
or, in the limit for which $r_{_0}$ approaches $r_{_H}\equiv 2\,M$ 
from outside,
\be
M\gg\ell_p\ \ \Leftrightarrow\ \  M_{_G}\gg m_p
\ .
\label{cond}
\ee
This coincides with what one would expect from 
a naive quantum mechanical argument:
if $\kappa\,\hbar/M$ is the Compton
wave length of the black hole, it should be much less
than its Schwarzschild radius $2\,M$ for the semiclassical
approximation to work.
\par
The remaining consistency conditions are that the R.H.S.s
of Eqs.~(\ref{wdw_g}) and (\ref{wdw_m}) be much smaller than
$\expec{\hat H_{_M}}$ (for fluctuations in the number of dust 
particles to be negligible)
and that in Eq.~(\ref{wdw_m}) the second term on
the L.H.S. be larger than the third (for the adiabatic approximation
to hold).
For $N\gg 1$, the above three conditions are essentially the same and,
on using Eq.~(\ref{m=sum}), one obtains the following result \cite{cv}
\be
M\gg {1/\mu}\ \ \Leftrightarrow\ \
M_{_G}\gg {{m_p}^2\over m}
\ ,
\label{cond_m}
\ee
which means that the Schwarzschild radius of the dust must
be much larger than the Compton wave length of the dust
particles.
Since $m\ll m_p$ for all known elementary particles, 
the latter expression is a stronger condition than 
Eq.~(\ref{cond}) so that, when it holds, one can safely
match the interior with an external classical space--time.
The fact that the adiabatic approximation is not valid
if Eq.~(\ref{cond_m}) is not satisfied is not surprising.
Indeed in such a case fluctuations, corresponding
to the creation of matter particles, are large \cite{cv}
(and the evaporation time small \cite{hawking}).
\section{Hawking radiation}
\label{radiation}
Let us consider the outer dust shell of the sphere situated at 
$\rho_{_0}$.
From the point of view of a static distant observer
the position of the shell is not a classical variable but is 
a quantum observable determined by the gravitational wave function 
describing the semiclassical collapse obtained in the previous 
section, Eq.~(\ref{psi}).
\par
We consider an isotropic massless scalar 
field $\varphi=\varphi(\rho,\eta)$ conformally coupled
to gravity \cite{birrell} and to a static emitter.
Its Lagrangian density will be given by
\begin{eqnarray}
{\cal L}_\varphi&=&-{1\over 2}\,\left[
\partial_\mu\varphi\,\partial^\mu\varphi
+{1\over 6}\,R\,\varphi^2\right]
\nonumber \\
& &+\int d\tau\,\int d\rho\,\int_0^{+\infty} dK\,
{|K\tilde\psi(K,\tau)|^2\over\bra{\tilde\psi}\,\hat K^2\,\ket{\tilde\psi}}
\nonumber \\
& & \times
\delta(\rho-\rho_{_0})\,
Q(\rho,\tau)\,\varphi(\rho,\tau)
\ ,
\end{eqnarray}
where $R$ is the curvature scalar, $Q(\rho,\tau)$ describes a particle 
(monopole) emitter and the factor $\delta(\rho-\rho_{_0})\,
|K\tilde\psi|^2/\bra{\tilde\psi}\,\hat K^2\,\ket{\tilde\psi}$
forces the interaction to be localized on the outer shell.
\par
Let us assume the emitter has a discrete set of internal
energy eigenstates described by $\ket{E}$
where $E$ is the energy observed by our distant observer.
We suppose the emitter is initially in an excited state
$E_{_0}$ and decays by emitting quanta of the scalar field
$\varphi$ to a state $E_{_0}-\hbar\,\omega$.
One may then estimate, using first order perturbation
theory, the total probability amplitude $P(\omega,\bar\eta)$
for the emitter to decay in a finite conformal time.
It will be given by
\begin{eqnarray}
P(\omega,\bar\eta)&=&{{Q_\omega}^2\over\hbar^2}\,
\int_{\tau(\eta_{_0})}^{\tau(\bar\eta)} d\tau''\,
\int_{\tau(\eta_{_0})}^{\tau(\bar\eta)} d\tau'  \nonumber \\
& & \times e^{i\,\omega\,(t''-t')}\,
D_\epsilon^+(\eta'',\eta';\rho)
\ ,
\label{p}
\end{eqnarray}
where
$Q_\omega\equiv|\bra{E_0}\,Q(\rho_{_0},0)\,\ket{E_0-\hbar\,\omega}|$
is the absolute value of the matrix
element of the monopole between the states with 
energy $E_0$ and $E_0-\hbar\,\omega$,
$t\simeq-2\,M\,\ln(\eta-\eta_{_H})$ is the Schwarzschild time measured 
by the static distant observer expressed in term of the conformal time,
$r(\eta_{_H})=r_{_H}$ and $\bar\eta$ is an upper cut off such that 
$\eta_{_H}<\bar\eta<\eta_{_0}$.
Further during the interval $(\eta_{_0},\bar\eta)$ the static
monopole emitter near the horizon is immersed in the gravitational
wave packet associated with the last dust shell and
$D^+_\epsilon$ is the Wightman function for the isotropic conformal 
scalar field in the Robertson--Walker metric \cite{birrell} 
evaluated at the same spatial point $\rho''=\rho'=\rho$, but with the 
$K_c$ dependence ``smeared'' by the wave function $K\,\tilde\psi$.
\par
One obtains a probability amplitude per unit (Schwarzschild) time equal 
to \cite{cv}
\be
\lim\limits_{T\to+\infty}\,
{P(\omega,T)\over T}\simeq{{Q_\omega}^2\,R^2(\infty)
\over 2\,\pi\,\hbar}\,{\omega\over 1-e^{-\beta\,\hbar\,\omega}}
\ ,
\label{p4}
\ee
which is a Planck distribution with the usual Hawking
temperature $T_{_H}=1/\beta=\hbar/8\,\pi\,M$ and
$R(\infty)$ is a non zero constant given by $R(\eta_{_H})$
with
\begin{eqnarray}
R(\eta)&\equiv&K_c(\eta)\,
{\bra{\tilde\psi(\eta)}\,\hat K\,\ket{\tilde\psi(\eta)}
\over\bra{\tilde\psi(\eta)}\,\hat K^2\,\ket{\tilde\psi(\eta)}}
\ ,
\label{r}
\end{eqnarray}
(see Ref.~\cite{cv} for explicit expressions).
\par
In the classical point--like limit, Eq.~(\ref{hp=0}),
$R\to 1$ and one recovers the usual field theory in the fixed 
Robertson--Walker--like background \cite{hawking,birrell}.
However, in alternative limit, Eq.~(\ref{b=0}),
one finds that the emitter decouples
from the conformal field, since $R\sim\alpha\to 0$,
and there is no emission within any finite time $T$,
\be
P(\omega,T)\to 0
\ .
\label{p=0}
\ee
One may speculate that this effect can be used to eliminate
ultra--planckian effects.
In fact, it is known that, if $\omega$ is the frequency
of the emitted quanta as is measured by a distant
observer, a fixed observer located near the point of emission
at $r=r_{_0}$ will measure instead a blue--shifted frequency
$\omega^\ast=(1-{2\,M/r_{_0}})^{-1/2}\,\omega$
which gives $\omega^\ast>m_p/\hbar={\ell_p}^{-1}$
for $r_{_0}$ sufficiently close to $r_{_H}$.
In order to create these modes, one must use an emitter
localized in a region smaller than $\omega^{-1}\sim\ell_p$. 
So one expects that $b$ in the wave function 
Eq.~(\ref{psi}) should be less than $\ell_p$
for our collapsing shell to couple with conformal 
quanta of ultra--planckian energies and
this would correspond to the limit in Eq.~(\ref{b=0})
which in turn implies Eq.~(\ref{p=0}).
\par
A further point worth noting is that $R(\infty)$ actually
depends on $\epsilon$ through $\alpha$ (see Eq.~(\ref{alpha})).
This of course implies that the probability amplitude
in Eq.~(\ref{p4}) depends on the internal dust sphere metric
and one has different emission rates depending on
the internal geometry.
\par
Let us end with a speculation:
we considered the matching condition in the classical limit;
if one wished to consider quantum mechanical corrections
one should replace it by the continuity of the gravitational
wave function inside the dust (Robertson--Walker)
with that outside (Schwarzschild).
This suggests the results in Eq.~(\ref{p4}) should not
change dramatically immediately outside the dust,
implying a form of quantum hair leading to information
on the geometry in the black hole through the intensity
of the radiation.

\end{document}